\documentstyle[preprint,aps,epsfig,psfig]{revtex}
\begin{document}

\draft

\title{Infrared Divergence and Twist-3 Distribution Amplitudes in QCD
Factorization For $B \rightarrow PP$
\footnote{Supported in part by
National Natural Science Foundation of China and State Commission of
Science and Technology of China}}

 \vspace{2cm}

\author{ Dongsheng Du${}^{1,2}$, Deshan Yang${}^{2}$ and Guohuai
Zhu${}^{2}$ \footnote{Email: duds@mail.ihep.ac.cn,
yangds@mail.ihep.ac.cn, zhugh@mail.ihep.ac.cn} } \address{${}^1$
CCAST
(World Laboratory), P.O.Box 8730, Beijing 100080, China\\ ${}^2$ Institute
of High Energy Physics, Chinese Academy of Sciences,
 P.O.Box 918(4), Beijing 100039, China
 \footnote{Mailing address}}

\date{\today}

\maketitle

\begin{abstract}
\indent
\tighten
Since b quark mass is not asymptotically large, chirally enhanced
corrections which arise from twist-3 wave functions may be important in B
decays. We thus evaluate the hadronic matrix elements with the final light
pseudoscalar
mesons described by leading twist and twist-3 distribution
amplitudes. We find that chirally enhanced corrections can be included
consistently in the framework of QCD factorization 
only if the twist-3 distribution amplitudes are symmetric. We then
give explicit expressions of $a_i^p$ for $B \rightarrow \pi\pi$ at the
next-to-leading order of
$\alpha_s$ including chirally enhanced corrections. We also briefly
discuss the divergence appeared in the hard spectator contributions.

\end{abstract}

\vspace{1.5cm}

{\bf PACS numbers 13.25.Hw 12.38.Bx}

\newpage

\narrowtext 
\tighten

Non-leptonic two-body B decays are crucial for us to extract CKM
matrix elements and uncover the origin of CP violations.
Experimentally, with the 
running of B factories, there will accumulate a great amount of data on
various B decay channels. Theoretically, however,
how to extract CKM matrix elements
from non-leptonic B rare decays with model-independence is still
an open question due to the complexity of strong interaction. In the
following, we first give a theoretical sketch on non-leptonic
B decays.

It is well known that the amplitude for the
decay $B \rightarrow P_1 P_2$ can be expressed as \cite{Buras}:
\begin{equation}
{\cal A}(B \rightarrow P_1 P_2) \propto \sum_i \lambda_i
C_i(\mu)
\langle P_1 P_2 \vert Q_i(\mu) \vert B \rangle,
\end{equation}
where $\lambda_i$ is a CKM factor, $C_i(\mu)$ is a Wilson coefficient
which
incorporates short distance contributions from strong interactions and
therefore is computable by making use of operator product expansion and 
renormalization group equations,
$\langle P_1 P_2 \vert Q_i(\mu) \vert B \rangle$
is a hadronic matrix element.  Obviously, if we want to extract CKM factor
from B decays, the hadronic matrix elements should be evaluated reliably.
However, due to our ignorance on hadronization, it would be a great
challenge to
calculate these hadronic matrix elements reliably from first principles. 
A commonly used approximation is naive factorization assumption, which is
based on Bjorken's color transparency argument \cite{Bjorken}: b quark
decays and
transfers a large momentum to final light quarks, in which two
fast-moving, nearly collinear 
final quarks with appropriate color can be viewed as a small color dipole
which will not significantly interact with the soft gluons
and finally form an emitted meson. Then we have:
\begin{equation}
\langle P_1 P_2 \vert Q_i(\mu) \vert B \rangle=
\langle P_1 \vert J_1 \vert 0 \rangle
\langle P_2 \vert J_2 \vert B \rangle,
\end{equation}
where $P_1$ labels the emitted meson and $P_2$ labels another light meson
which absorbs the spectator quark from B meson.
This approximation completely ignores non-factorizable
contributions which connect the emitted meson to the spectator system and
expresses the hadronic matrix
elements in terms of
meson decay constants and form factors. Since decay constants and form
factors can be, at least in principle, well determined from other
experiments, the branching ratios of non-leptonic B decays are obtained
under this assumption. The main deficiency of this approximation is that
non-factorizable contributions are completely missing. In
consequence, the hadronic matrix elements lose their scheme- and
scale-dependence.
Noting that Wilson coefficients are scheme- and scale-dependent, the
corresponding decay width will also depend on renormalization scheme and
scale which is
unphysical. This is a clear indication that non-factorizable
contributions, which amount to final-state rescattering and strong
interaction phase shift, are important. Several generalizations of naive
factorization assumption
have been proposed to phenomenologically parameterize non-factorizable
contributions. Since this kind of parameterization has no relation to, and
therefore does not gain any information from, the underlying QCD dynamics,
the resulting predictions on B decays are still model-dependent.

In ref \cite{BBNS1,BBNS2}, Beneke, Buchalla, Neubert and Sachrajda
proposed a
promising QCD factorization method: The hadronic matrix elements 
$\langle P_1 P_2 \vert Q_i(\mu) \vert B \rangle$ contain two distinct
scale: one is a
large scale $\mu={\cal O}(m_b)$, the other is $\Lambda_{QCD}$ which is
the scale of hadronization. In the heavy quark limit, they show that    
the short distance contributions which are related to the large scale
$\mu={\cal O}(m_b)$ can be, at least at one-loop order, separated from
the long distance effects and
are thus calculable. Furthermore, the long distance effects can be
parameterized by light-cone distribution amplitudes and non-perturbative
form factors. Thus, the factorization formula can be explicitly expressed
as:
\cite{BBNS1,BBNS2}
\begin{equation}
\langle P_1 P_2 \vert Q_i \vert B \rangle =
F^{B \rightarrow P_2}(0) \int \limits_0^1 dx T^I_i(x) \Phi_{P_1}(x)
+\int \limits_0^1 d\xi dx dy T_i^{II}(\xi,x,y) \Phi_B(\xi) \Phi_{P_1}(x)
\Phi_{P_2}(y),
\end{equation}
where $\Phi_B(\xi)$ and $\Phi_{P_i}(x)$ are the
leading-twist light-cone distribution amplitudes of B and the final light 
mesons     
respectively, $T^{I,II}_i$ denote hard-scattering kernels which are
calculable order by order in perturbative theory. This formula holds for 
the case that the emitted meson $P_1$ is a light meson
\cite{BBNS1,our,yyd} or an onia of two heavy quarks 
\cite{BBNS2,Cheng,Chay} no matter whether
$P_2$ is a heavy or light meson. But in
this article we will focus on the case that B decays to two light 
pseudoscalar mesons.

In ref \cite{BBNS1,BBNS2}, the authors pointed out that the equality
sign of 
eq. (3) is valid only in the heavy quark limit. So if the heavy quark
limit is an adequate approximation for B meson, or in another word, if
power corrections in $1/m_b$ can be safely neglected, then everything is
perfect. At the zero order of $\alpha_s$, it can reproduce "naive
factorization", at the higher order of $\alpha_s$, the corrections can
be systematically calculated in Perturbative QCD which will restore the 
scheme- and scale- dependence for the hadronic matrix elements. Therefore, 
the decay amplitudes of B meson can be reliably evaluated from first
principles, and the necessary inputs are heavy-to-light form factors and
light-cone distribution amplitudes. But in the real world, bottom quark
mass is not asymptotically large(but about $4.8~GeV$), therefore it may be
necessary to consider power corrections in $1/m_b$. Unfortunately there
are a variety of sources which may contribute to power corrections in
$1/m_b$, examples are higher twist distribution amplitudes, hard spectator
interaction and transverse momenta of quarks in the light meson.
Furthermore, there is no known systematic way to evaluate these
power corrections for exclusive decays.
Though naively, it is expected that power corrections may be neglected
because $\Lambda_{QCD}/m_b \simeq 1/15$ is a small number, power
suppression may numerically fail in some cases.  An obvious and possibly
the most important case
is chirally enhanced power corrections. As pointed out in ref
\cite{BBNS1},
numerically the enhanced factor
$r_{\chi}=\frac{2 m_{\pi}^2}{m_b(m_u+m_d)} \simeq 1.18$ which makes
the power suppression completely fail. This parameter is multiplied
by $a_6$ and $a_8$, where $a_6$ is very important numerically in
penguin-dominated B decays. So an evaluation of the hadronic matrix
elements including chirally enhanced corrections may be
phenomenologically or numerically important. In the following, we will
examine this problem in some details.
 
Chirally enhanced corrections arise from twist-3 light-cone distribution
amplitudes, generally called $\Phi_p(x)$ and $\Phi_{\sigma}(x)$. For light
pseudoscalar mesons, they are defined as \cite{Braun}
\begin{eqnarray}
\langle P(p') \vert {\bar q}(y) {\it i} \gamma_5 q(x) \vert 0 \rangle
&=& f_P \mu_P \int_0^1 {\it du~e}^{i(up' \cdot y + {\bar u}p' \cdot x)}
\Phi_p(u), \\
\langle P(p') \vert {\bar q}(y) \sigma_{\mu \nu} \gamma_5 q(x) \vert 0
\rangle &=& if_P \mu_P (p^{\prime}_{\mu} z_{\nu} - p^{\prime}_{\nu}
z_{\mu} ) \int_0^1 {\it du~e}^{i(up' \cdot y + \bar u p' \cdot x)}
\frac{\Phi_{\sigma}(u)}{6},
\end{eqnarray}
where $\mu_{p}=\frac{M^2_{p}}{m_1+m_2}$, $z=y-x$,
$m_1$ and $m_2$ are the corresponding current quark masses. 
If we want to generalize 
QCD factorization method to include chirally enhanced corrections
consistently, we should describe the emitted light meson with leading
twist-2 and twist-3 distribution amplitudes \cite{Beneke2}:
\begin{eqnarray}
\langle P(p') \vert {\bar q_{\alpha}}(y) q_{\delta}(x) \vert 0 \rangle
&=&\frac{if_P}{4} \int_0^1{\it du~e}^{i(up' \cdot y + {\bar u}p' \cdot
x)} \nonumber \\
&\times& \left \{ \slash{\hskip -2.5mm}p^{\prime} \gamma_5 \Phi(u)
-\mu_P \gamma_5 \left ( \Phi_p(u)-\sigma_{\mu \nu}p^{\prime \mu}
z^{\nu} \frac{\Phi_{\sigma}(u)}{6} \right ) \right \}_{\delta \alpha}.
\end{eqnarray}
A technical proof 
of factorization requires that the hard scattering kernels in Eq.(3) are
infrared finite.
Authors of Ref \cite{BBNS1,BBNS2} have shown it explicitly with
leading twist distribution amplitudes.
Then a basic and perhaps a
difficult task for us is to show the infrared finiteness of the 
hard-scattering kernels using twist-3
distribution amplitudes after summing over the four vertex correction
diagrams (Fig. 1(a)-(d)).

The start point for B decays is $\vert\Delta B\vert=1$ effective
Hamiltonian \cite{Buras}:
\begin{eqnarray}
{\cal{H}}_{eff}&=& \frac{G_F}{\sqrt{2}}
 \left [ \sum_{q=u,c} v_q \left( C_1(\mu) Q^q_1(\mu)+ C_2(\mu)Q^q_2(\mu)
  + \sum_{k=3}^{10} C_k(\mu)Q_k(\mu)  \right) \right. \nonumber \\
&& - \left. v_t(C_{7\gamma}Q_{7\gamma}+C_{8G}Q_{8G}) \frac{}{} 
\right ]+h.c.,
\end{eqnarray}
where $v_q=V_{qb}V_{qd}^{*}$(for $b\rightarrow d$ transition) or
 $v_q=V_{qb}V_{qs}^{*}$(for $b\rightarrow s$ transition) and
$C_i(\mu)$ are Wilson coefficients which have been evaluated to
next-to-leading order approximation. The four-quark operators $Q_i$ are
\begin{equation}
\begin{array}{l}
\begin{array}{ll}
Q^u_1= ( \bar{u}_{\alpha} b_{\alpha} )_{V-A}
         ( \bar{q}_{\beta} u_{\beta} )_{V-A}&
Q^c_1= ( \bar{c}_{\alpha} b_{\alpha} )_{V-A}
         ( \bar{q}_{\beta} c_{\beta} )_{V-A}\\
Q^u_2= ( \bar{u}_{\alpha} b_{\beta} )_{V-A}
         ( \bar{q}_{\beta} u_{\alpha} )_{V-A}&
Q^c_2= ( \bar{c}_{\alpha} b_{\beta} )_{V-A}
         ( \bar{q}_{\beta} c_{\alpha} )_{V-A}\\
Q_3= (\bar{q}_{\alpha} b_{\alpha} )_{V-A}
      \sum\limits_{q'}
     ( \bar{q}^{'}_{\beta} q^{'}_{\beta} )_{V-A}&
Q_4= (\bar{q}_{\beta} b_{\alpha} )_{V-A}
      \sum\limits_{q'}
     ( \bar{q}^{'}_{\alpha} q^{'}_{\beta} )_{V-A}\\
Q_5= (\bar{q}_{\alpha} b_{\alpha} )_{V-A}  
      \sum\limits_{q'}
      ( \bar{q}^{'}_{\beta} q^{'}_{\beta} )_{V+A}&
Q_6= (\bar{q}_{\beta} b_{\alpha} )_{V-A}
      \sum\limits_{q'}
     ( \bar{q}^{'}_{\alpha} q^{'}_{\beta} )_{V+A}\\
Q_7= \frac{3}{2} (\bar{q}_{\alpha} b_{\alpha} )_{V-A}
      \sum\limits_{q'} e_{q'}
     ( \bar{q}^{'}_{\beta} q^{'}_{\beta} )_{V+A}&
Q_8=\frac{3}{2}  (\bar{q}_{\beta} b_{\alpha} )_{V-A}
   \sum\limits_{q'} e_{q'}
    ( \bar{q}^{'}_{\alpha} q^{'}_{\beta} )_{V+A}\\
Q_9= \frac{3}{2} (\bar{q}_{\alpha} b_{\alpha} )_{V-A}
      \sum\limits_{q'} e_{q'}
    ( \bar{q}^{'}_{\beta} q^{'}_{\beta} )_{V-A}&
Q_{10}=\frac{3}{2}  (\bar{q}_{\beta} b_{\alpha} )_{V-A}
      \sum\limits_{q'} e_{q'}
     ( \bar{q}^{'}_{\alpha} q^{'}_{\beta})_{V-A}\\
\end{array}
\end{array}
\end{equation}
and
\begin{equation}
Q_{7\gamma}=\frac{e}{8\pi^2} m_b \bar{q}_{\alpha} \sigma^{\mu\nu}
(1+\gamma_5) b_{\alpha} F_{\mu\nu}, ~~
Q_{8G}=\frac{g}{8\pi^2} m_b \bar{q}_{\alpha} \sigma^{\mu\nu}
t^{a}_{\alpha \beta} b_{\beta} G^a_{\mu\nu}, ~~(q=d~ {\rm or} ~s).
\end{equation}
With these effective operators,
$B \rightarrow P_1 P_2$ decay amplitudes
in QCD factorization can be written as:
\begin{equation}
A(B\rightarrow P_1 P_2)=\frac{G_F}{\sqrt{2}}
\sum \limits_{p=u,c} \sum \limits_{i=1,10} v_p a^p_i
\langle P_1 P_2 \vert Q_i \vert B \rangle_F,
\end{equation}
where $v_p$ is CKM factor,
$\langle P_1 P_2 \vert Q_i \vert B \rangle_F$ is the factorized   
matrix element. We will calculate QCD coefficients $a_i^p$
and show explicitly that they are infrared finite.

Infrared divergences exist in vertex correction diagrams (Fig.1(a)-(d)),
so
let us first consider these diagrams. For $(V-A) \bigotimes (V-A)$ and
$(S+P) \bigotimes (S-P)$ operators, twist-3 distribution amplitudes make
no contribution because of their Lorentz structures. Therefore, QCD
coefficients $a_i^p$ (except for $a_6^p$ and $a_8^p$)
are nearly as same as those 
obtained in Ref \cite{BBNS1,our,yyd}
where leading twist distribution amplitudes are considered.
The only difference is hard-spectator term (Fig.1(g)-(h)) which have been
shown in Ref
\cite{BBNS3,Beneke}, we will discuss it later. As to $(V+A) \bigotimes
(V-A)$ operator, there are some subtleties in regularizing the infrared
divergences. If we use dimension regularization, the infrared finiteness
will not hold after summing over those four vertex correction diagrams.
That is because wave functions are defined in 4-dimensions, it may be
unconsistent to naively extend its usage to d-dimensions. Thus we assign a
virtual mass to the gluon propagator and regularize the infrared integrals
in four dimensions. For the twist-3 distribution amplitudes $\Phi_p(x)$,
the calculations are performed in momentum space. Then it is 
straightforward to verify that the vertex correction contributions of
$(V+A) \bigotimes (V-A)$ operator to $(S+P) \bigotimes (S-P)$ are
infrared finite:
\begin{equation}
V = -\frac{\alpha_s}{4 \pi}\frac{C_F}{N}
 \int_0^1 dx \Phi_p(x) \{
i \pi \log \frac{x}{\bar x} + \log \frac{x}{\bar x} 
-Li_2(-\frac{\bar x}{x})
+Li_2(-\frac{x}{\bar x})+6 \},
\end{equation}
where $\bar x = 1-x$ and $Li_2(x)$ is dilogarithm function.
On the other hand, when considering $\Phi_{\sigma}$, we have to do
the calculations in coordinate space
according to Eq.(5). For example, let us consider Fig. 2. In 
coordinate space, we have:
\begin{eqnarray}
{\cal A} &=& \int d^4 x_1 d^4 x_2 d^4 x_3 
~{\bar u_{\rho}}e^{iq^{\prime} \cdot x_3}~
\gamma_{\lambda}(1+\gamma_5) 
\int \frac{d^4 k_2}{(2 \pi)^4}\frac{i \slash{\hskip -2.5mm}k_2}{k_2^2}
e^{-i k_2 \cdot (x_3-x_2)}~
[ -i g_s \gamma^{\alpha} (T^a)_{\rho \sigma} ] \nonumber \\
& &\frac{i f_P\mu_P}{4 N_c} \sigma_{\mu \nu} \gamma_5
q^{\mu}(x_3-x_2)^{\nu}
\int dv e^{i (v q \cdot x_3 + {\bar v}q \cdot x_2)}
\frac{\Phi_{\sigma}(v)}{6}\delta_{\beta \sigma}~
\gamma^{\lambda}(1-\gamma_5) \nonumber \\
& & \int \frac{d^4 k_1}{(2 \pi)^4}
\frac{i(\slash{\hskip -2.5mm}k_1+m_b)}{k_1^2-m_b^2}
e^{-i k_1 \cdot (x_3-x_1)}~[-i g_s \gamma_{\alpha} (T^a)_{\beta \alpha} ]~
e^{-i p \cdot x_1} b_{\alpha}~
\int \frac{d^4 k}{(2 \pi)^4}\frac{-i}{k^2-m_g^2} e^{-i k \cdot (x_2-x_1)} 
\nonumber \\
&=& \int dv d^4 x_1 d^4 x_2 d^4 x_3 \frac{d^4 k}{(2 \pi)^4}
\frac{d^4 k_1}{(2 \pi)^4} \frac{d^4 k_2}{(2 \pi)^4}~{\bar u_{\rho}}
e^{iq^{\prime} \cdot x_3}~e^{-i k_2 \cdot (x_3-x_2)}\nonumber \\ 
& &\frac{i f_P\mu_P}{4 N_c}
q^{\mu}(x_3-x_2)^{\nu} \frac{\Phi_{\sigma}(v)}{6} \delta_{\beta \sigma} 
e^{i (v q \cdot x_3 + {\bar v}q \cdot x_2)}~e^{-i k_1 \cdot (x_3-x_1)}
~e^{-i k \cdot (x_2-x_1)}~e^{-i p \cdot x_1} b_{\alpha} \cdot  
H_{\mu \nu}(k,k_1,k_2)\nonumber \\
&=& \left . (2 \pi)^4 \delta^4 (p-q-q^{\prime})~\frac{f_P \mu_P}{4 N_c}
\int dv \frac{\Phi_{\sigma}(v)}{6} 
\int \frac{d^4 k}{(2 \pi)^4}~
{\bar u_{\rho}} q^{\mu}\frac{\partial}{\partial k_{2 \nu}}
H_{\mu \nu}(k,p-k,k_2) \right |_{k_2=k-{\bar v}q}~b_{\alpha},   
\end{eqnarray}
where $H_{\mu\nu}(k,k_1,k_2)$ contains the Lorentz structure and
propagators of the hard scattering kernels:
\begin{eqnarray}
H_{\mu \nu}(k,k_1,k_2)&=&\gamma_{\lambda}(1+\gamma_5)~
\frac{i \slash{\hskip -2.5mm}k_2}{k_2^2}
[ -i g_s \gamma^{\alpha} (T^a)_{\rho \sigma} ]~
\sigma_{\mu \nu} \gamma_5~\gamma^{\lambda}(1-\gamma_5) \nonumber \\
&&
\frac{i (\slash{\hskip -2.5mm}k_1+m_b)}{k_1^2-m_b^2}~
[ -i g_s \gamma_{\alpha} (T^a)_{\beta \alpha}
]~\frac{-i}{k^2-m_g^2}\delta_{\beta \sigma}.
\end{eqnarray}
After a lengthy derivation, we can regularize the infrared divergences
with a gluon virtual mass $m_g$:
\begin{eqnarray}
{\rm Fig.1(a)} &\sim& -\frac{\alpha_s}{4 \pi} \frac{C_F}{N_c}
\frac{\Phi_{\sigma}(v)}{v}
\{\frac{\log^2 \mu}{2}+2 \log(-v)\log\mu-4\log v\log\mu+\log\mu 
+{\rm finite~terms} \}~, \\
{\rm Fig.1(b)} &\sim& \frac{\alpha_s}{4 \pi} \frac{C_F}{N_c}
\frac{\Phi_{\sigma}(v)}{\bar v}
\{\frac{\log^2 \mu}{2}+2 \log(-{\bar v})\log\mu-4\log{\bar
v}\log\mu+\log\mu   
+{\rm finite~terms} \} ~,\\
{\rm Fig.1(c)} &\sim& \frac{\alpha_s}{4 \pi} \frac{C_F}{N_c}
\frac{\Phi_{\sigma}(v)}{v}
\{ \log^2 \mu -2 \log(-v)\log\mu + 3 \log\mu +{\rm finite~terms} \} ~.\\
{\rm Fig.1(d)} &\sim& -\frac{\alpha_s}{4 \pi} \frac{C_F}{N_c}
\frac{\Phi_{\sigma}(v)}{\bar v}
\{ \log^2 \mu -2 \log(-{\bar v})\log\mu + 3 \log\mu +{\rm finite~terms}
\}~,
\end{eqnarray}
where $\mu=m_g^2/m_b^2$. From the above equations, it is observed that,
in the case of $\Phi_{\sigma}$ distribution amplitudes, the terms with
infrared divergence in vertex
correction diagrams can not cancel unless $\Phi_{\sigma}(v)$
is a symmetric function: $\Phi_{\sigma}(v)=\Phi_{\sigma}(\bar v)$.
This is an unexpected result, which means QCD factorization is violated 
for asymmetric twist-3 light-cone distribution amplitudes. This indicates
that chirally enhanced corrections can be included consistently in the
framework of QCD factorization only when twist-3 light-cone distribution
amplitudes are symmetric. Therefore, in the following, we will implicitly 
assume a symmetric  twist-3 light-cone distribution amplitude for light
pseudoscalar mesons. It is then straightforward to show that vertex
corrections of $(V+A) \otimes (V-A)$ operator are completely canceled
after summing over four diagrams in the case of $\Phi_{\sigma}$
distribution amplitude. 

For penguin contractions (Fig.1(e)-(f)) and hard spectator diagrams
(Fig.1(g)-(h)), we shall also do the calculations in coordinate space 
when $\Phi_{\sigma}(v)$ is included. When treating penguin contractions,
it should be careful that Fig.1(e) contains two kinds of topology, which
is
displayed in Fig.3. They are equivalent in 4 dimensions according to Fierz 
relations. However, since penguin corrections contain ultraviolet
divergences, we must do calculations in d dimensions where these two
kinds of topology are not equivalent \cite{Buras2}. We did not notice it
and therefore obtained a wrong term $-\frac{2f}{3}C_4$ in
the expression of $a_4^p$ in \cite{our}. We also obtained a wrong term
$(C3+C4/N)/3$ and missed a term of $(C4+C3/N)$ in the expression of 
$a_{10}^p$ in \cite{our} for the same reason.

 Then as an illustration, the explicit expressions of $a_i^p$
($i=1$ to $10$) for $B \rightarrow \pi\pi$ (using symmetric
light-cone distribution amplitudes of the pion) are 
obtained. But it is easy to generalize these formulas to 
the case that the final states are other light pseudoscalars.
We now list $a_i^p$ for $B \rightarrow \pi \pi$ as follows: 
\begin{eqnarray}
a_1^u&=&C_1+\frac{C_2}{N} + \frac{\alpha_s}{4\pi} \frac{C_F}{N} C_2 F, \\ 
a_2^u&=&C_2+\frac{C_1}{N} + \frac{\alpha_s}{4\pi} \frac{C_F}{N} C_1 F,\\
a_3&=&C_3+\frac{C_4}{N} + \frac{\alpha_s}{4\pi} \frac{C_F}{N} C_4 F, \\
a_4^p&=&C_4+\frac{C_3}{N} + \frac{\alpha_s}{4\pi} \frac{C_F}{N} C_3 F
\nonumber \\
& &- \frac{\alpha_s}{4\pi} \frac{C_F}{N} \left \{
C_1 (\frac{4}{3}\log\frac{\mu}{m_b}+G(s_p)-\frac{2}{3})+
(C_3-\frac{C_9}{2})(\frac{8}{3}\log\frac{\mu}{m_b}+G(0)+G(1)-\frac{4}{3}) 
\right. \nonumber \\
& & +\sum_{q=u,d,s,c,b}
(C_4+C_6+\frac{3}{2}{\rm e_q}C_8+\frac{3}{2}{\rm e_q}C_{10}) \left. 
(\frac{4}{3}\log\frac{\mu}{m_b}+G(s_q))+G_8 C_{8G} \right \}, \\
a_5&=&C_5+\frac{C_6}{N}+\frac{\alpha_s}{4\pi}\frac{C_F}{N} C_6(-F-12),\\
a_6^p&=&C_6+\frac{C_5}{N} -\frac{\alpha_s}{4\pi} \frac{C_F}{N}6C_5 
\nonumber \\
& &-\frac{\alpha_s}{4\pi} \frac{C_F}{N} \left \{
C_1 ((1+\frac{2}{3}A_{\sigma})\log\frac{\mu}{m_b}-\frac{7}{12}-
\frac{1}{2}A_{\sigma}+G^{\prime}(s_p)+G^{\sigma}(s_p)) 
\right. \nonumber \\
& &+
\sum_{q=d,b}(C_3-\frac{C_9}{2})
((1+\frac{2}{3}A_{\sigma})\log\frac{\mu}{m_b}-\frac{7}{12}-
\frac{1}{2}A_{\sigma}+G^{\prime}(s_q)+G^{\sigma}(s_q)) \nonumber \\
& &+\sum_{q=u,d,s,c,b} 
(C_4+C_6+\frac{3}{2}{\rm e_q}C_8+\frac{3}{2}{\rm e_q}C_{10})
\left( (1+\frac{2}{3}A_{\sigma})\log\frac{\mu}{m_b}-\frac{1}{12}-
\frac{1}{6}A_{\sigma}+G^{\prime}(s_q)+G^{\sigma}(s_q) \right) \nonumber \\
& & \left. +(\frac{3}{2}+A_{\sigma})C_{8G} \right \}, \\
a_7&=&C_7+\frac{C_8}{N}+\frac{\alpha_s}{4\pi}\frac{C_F}{N} C_8(-F-12), \\
a_8^p&=&C_8+\frac{C_7}{N} -\frac{\alpha_s}{4\pi} \frac{C_F}{N}6C_7  
\nonumber \\
& &-\frac{\alpha_{em}}{9\pi} \left \{
(C_2+\frac{C_1}{N}) 
((1+\frac{2}{3}A_{\sigma})\log\frac{\mu}{m_b}-\frac{7}{12}-
\frac{1}{2}A_{\sigma}+G^{\prime}(s_p)+G^{\sigma}(s_p))
\right. \nonumber \\
& &+
(C_4+\frac{C_3}{N}) \sum_{q=d,b} \frac{3}{2}{\rm e_q}
((1+\frac{2}{3}A_{\sigma})\log\frac{\mu}{m_b}-\frac{7}{12}-
\frac{1}{2}A_{\sigma}+G^{\prime}(s_q)+G^{\sigma}(s_q)) \nonumber \\
& &+(C_3+\frac{C_4}{N}+C_5+\frac{C_6}{N}) \sum_{q=u,d,s,c,b} 
\frac{3}{2}{\rm e_q}
\left( (1+\frac{2}{3}A_{\sigma})\log\frac{\mu}{m_b}-\frac{1}{12}-
\frac{1}{6}A_{\sigma}+G^{\prime}(s_q)+G^{\sigma}(s_q) \right) \nonumber \\
& & \left. 
+(\frac{3}{4}+\frac{1}{2}A_{\sigma})C_{7\gamma} \right \}, \\
a_9&=&C_9+\frac{C_{10}}{N}+\frac{\alpha_s}{4\pi} \frac{C_F}{N} C_{10} F, \\
a_{10}^{p}&=&C_{10}+\frac{C_9}{N}+
\frac{\alpha_s}{4\pi} \frac{C_F}{N}C_{9} F
- \frac{\alpha_{em}}{9\pi} \left \{
(C_2+\frac{C_1}{N}) (\frac{4}{3}\log\frac{\mu}{m_b}+G(s_p)-\frac{2}{3})
\right. \nonumber \\
& &+(C_4+\frac{C_3}{N})\sum_{q=d,b} \frac{3}{2}{\rm e_q} 
(\frac{4}{3}\log\frac{\mu}{m_b}+G(s_q)-\frac{2}{3}) \nonumber \\
& & +(C_3+\frac{C_4}{N}+C_5+\frac{C_6}{N}) \sum_{q=u,d,s,c,b} \left.
\frac{3}{2}{\rm e_q}
(\frac{4}{3}\log\frac{\mu}{m_b}+G(s_q))+\frac{1}{2}G_8 C_{7\gamma} 
\right \}.   
\end{eqnarray}
Here $N=3$ is the number of color,
$C_F=\frac{N^2-1}{2N}$ is the factor of color,
$s_q=m_q^2/m_b^2$ and we define the other symbols
in the above expressions as:
\begin{eqnarray}
&&F=-12 \ln \frac{\mu}{m_b} -18+f^{I}+f^{II}, \\
&&f^{I}=\int \limits_{0}^{1}~ dx~g(x)\Phi(x),
~G_8=\int \limits_{0}^{1}~ dx~G_8(x) \Phi(x), \\
&&G(s)=\int \limits_{0}^{1}~ dx~G(s,x) \Phi(x), \\
&&G^{\prime}(s)=\int \limits_{0}^{1}~ dx~G^{\prime}(s,x) \Phi_p(x),\\
&&G^{\sigma}(s)=\int \limits_{0}^{1}~ dx~G^{\sigma}(s,x) 
\frac{\Phi_{\sigma}(x)}{6(1-x)},~~~~~
A_{\sigma}=\int \limits_{0}^{1}~ dx~ \frac{\Phi_{\sigma}(x)}{6(1-x)},
\end{eqnarray}
here $\Phi(x)$($\Phi_p(x)$,$\Phi_{\sigma}(x)$) is leading twist (twist-3)
wave
function of the emitted pion, and the hard-scattering functions are
\begin{eqnarray}
&&g(x)=3 \frac{1-2x}{1-x} \ln x - 3 i \pi, ~~G_8(x)=\frac{2}{1-x}, \\
&&G(s,x)=-4 \int \limits_{0}^{1}~ du~u(1-u) \ln (s-u(1-u)(1-x)-i
\epsilon), \\
&&G^{\prime}(s,x)=-3 \int \limits_{0}^{1}~ du~u(1-u) \ln (s-u(1-u)(1-x)-i
\epsilon), \\
&&G^{\sigma}(s,x)=-2 \int \limits_{0}^{1}~ du~u(1-u) \ln (s-u(1-u)(1-x)-i
\epsilon) \nonumber \\
&&~~~~~~~~~~~~~~~
+ \int \limits_{0}^{1}~ du~ \frac{u^2(1-u)^2(1-x)}{s-u(1-u)(1-x)-i \epsilon}.
\end{eqnarray}
The contributions from the hard spectator scattering (Fig.1(g)-(h))
are reduced into the factor $f^{II}$. We take the wave function of B meson
as $\gamma_5 (\slash{\hskip -2.5mm}P_B -M_B )\Phi_B(\xi)$. Then an
explicit calculations show that twist-3 distribution amplitudes of the
emitted pion make no contributions to $f^{II}$. It means that
there is no hard spectator contributions for $a_6^p$ and $a_8^p$. For
other QCD coefficients $a_i^p$, we have:
 \begin{equation}
f^{II}=\frac{4 \pi^2}{N}
\frac {f_{\pi}f_B}{F^{B\rightarrow \pi}_{+}(0) m_B^2}
\int \limits_{0}^{1}~ d\xi~ \frac{\Phi_B(\xi)}{\xi}
\int \limits_{0}^{1}~ dx~ \frac{\Phi(x)}{x} \int \limits_{0}^{1}~
dy~ \left [ \frac{\Phi(y)}{1-y}+\frac{2 \mu_{\pi}}{M_B}
\frac{\Phi_{\sigma}(y)}{6(1-y)^2} \right ].
\end{equation}
Here $\Phi(x)$ is leading twist distribution amplitude of the emitted
pion, $\Phi(y)$($\Phi_{\sigma}(y)$) is twist-2(twist-3)
distribution amplitudes of the recoiled pion. This formula is consistent 
with the result of Ref. \cite{Cheng}.

In the above expressions of $a_i^p$, $a_6^p$ and $a_8^p$ can now be
evaluated to next-to-leading order of $\alpha_s$, which significantly
reduce their scale-dependence. As to other QCD coefficients $a_i^p$, there
contains a divergent integral in hard spectator term $f^{II}$. In the next
paragraph, we will argue that this disturbing divergence may need further
consideration. Here we simply assume that 
$\int \frac{dy}{y} \sim ln \frac{m_b}{\Lambda_{QCD}}$ (similar to
what
have been done in Ref \cite{BBNS3,Beneke}, though our assumption here is
certainly an oversimplification). We thus illustrate numerically the
scale-dependence of $a_i^p(\pi\pi)$ in Table.1. Here we use the
asymptotic distribution amplitudes 
\begin{equation}
\Phi(x)=\Phi_{\sigma}(x)=6x(1-x)~~{\rm and}~~
\Phi_p(x)=1,
\end{equation}
and the input parameters are taken as follows: 
$F^{B\pi}(0)=0.33$, $f_B=0.2~GeV$, $f_{\pi}=133~MeV$, 
the pole masses $m_b=4.8~GeV$, $m_c=1.4~GeV$, the $\overline {MS}$ masses
${\overline m_t}({\overline m_t})=170~GeV$, 
${\overline m_b}({\overline m_b})=4.4~GeV$,
$\overline {m}_u(2~GeV)=4.2~MeV$, $\overline{m}_d(2~GeV)=7.6~MeV$ and
$\Lambda_{QCD}^{(5)}=225~MeV$.

We notice that the above approach of evaluating hard spectator
contribution
is naive. For instance, the scale of hard spectator contribution should
be different from the vertex correction contribution. While it seems  
reasonable to take the scale
$\mu \sim {\cal O}(m_b)$ for the vertex correction diagrams to avoid large
logarithm
$\alpha_s \log \frac{\mu}{m_b}$, a natural choice  of the scale of hard
spectator contribution may be around ${\cal O}(1~GeV)$ because the
average
momentum squared of the exchanged gluon is about $1~ GeV^2$.
Another disturbing feature of hard spectator contribution is that, as 
pointed out in
ref \cite{BBNS3,Beneke}, when including the contribution of
$\Phi_{\sigma}$,   
there would appear divergent integral $\int_0^1 dy \frac{1}{y}$ even if
the symmetric distribution amplitude is applied. This divergent integral
implies
that the dominant contribution comes from the end-point region, or in   
another word, it is dominated by soft gluon exchange. However the  
transverse momentum may not be omitted in the end-point region
\cite{Huang}, if so, the
corresponding divergent integral would then changed to:
\begin{equation}
\int dy \frac{1}{y} \rightarrow
\int dy d^2 k_T \frac{\Psi(y,k_T)}{y \xi m_b^2 +k^2_T}.
\end{equation}
As an illustration, we do not consider the $k_T$ dependence of wave
functions (though it is certainly not a good approximation), then the   
above integral is proportional to:
\begin{equation}
\int \frac{dy d k^2_T}{y \xi m_b^2 +k^2_T} \propto
\int \frac{dx dy}{x+y}.
\end{equation}
The above integration converges now, furthermore it is not dominated
by end-point contribution. This illustrates that the   
treatment of hard spectator diagrams may need further
discussions. 

There exists "annihilation" contributions which may belong to
chirally enhanced corrections. In Ref. \cite{Beneke}, the authors have
discussed this topic and find that a divergent integral 
$(\int \frac{dx}{x})^2$ will appear. We suspect that this divergence    
may disappear, similar to the hard spectator term,  if
the effect of transverse momenta can be included.
It is also possible that "annihilation" contributions are really dominated 
by soft interactions and thus violate factorization. Due to its
complexity, we do not include "annihilation" contributions in the
expressions of $a_i^p$.  

In summary, to generalize QCD factorization method to include chirally
enhanced corrections consistently, the final light mesons should be described 
with leading twist and twist-3 distribution amplitudes. We demonstrate
that the infrared finiteness of the hard scattering kernels can be
obtained only if the twist-3 distribution amplitudes are symmetric. We
then
give explicit expressions of $a_i^p$ at next-to-leading order of 
$\alpha_s$ including chirally enhanced corrections. We also discuss
briefly the disturbing hard spectator contributions.

\section*{Acknowledgements}
We thank Prof. Hai-Yang Cheng for pointing out errors in the
coefficients of $C_{7\gamma}$ and $C_{8G}$
and Prof. Mao-Zhi Yang for helpful discussions.
This work is supported in part by National Natural
Science Foundation of China and State Commission of
Science and Technology of China.

\narrowtext
\tighten

\begin{table}
\vspace*{1cm}
\begin{tabular}{ccccc}
QCD &  \multicolumn{2}{c} {$\mu=5.0~GeV$} &
\multicolumn{2}{c} {$\mu=2.5~GeV$} \\  
Coefficients & NLO & LO  & NLO & LO \\ \hline
$a_1^u$ & $1.024+0.012 i$   & $1.017$  &
          $1.034+0.024 i$   & $1.037$  \\
$a_2^u$ & $0.144-0.076 i$ & $0.188$  &
          $0.123-0.100 i$ & $0.109$  \\ \hline
$a_3$   & $0.003+0.002 i$ & $0.002$  &
          $0.004+0.004  i$ & $0.004$   \\
$a_4^u$ & $-0.027-0.014 i$ & $-0.029$ &
          $-0.029-0.017 i$ & $-0.040$ \\   
$a_4^c$ & $-0.033-0.007 i$ & $-0.029$ &
          $-0.036-0.007 i$ & $-0.040$  \\
$a_5$   & $-0.003-0.003 i$ & $-0.005$ & 
          $-0.002-0.005 i$ & $-0.010$ \\ 
$r_{\chi} a_6^u$ & $-0.036-0.012 i$ & $-0.033$ &
          $-0.037-0.011 i$ & $-0.040$ \\
$r_{\chi} a_6^c$ & $-0.039-0.005 i$ & $-0.033$   &
          $-0.040-0.004 i$ & $-0.040$ \\ \hline
$a_7 \times 10^{5}$   & $11.9+2.8 i$     & $13.8$   &
          $0.0+5.4 i$     & $7.6$  \\      
$r_{\chi} a_8^u \times 10^{5}$ & $36.8-10.9 i$     & $36.8$   &
          $45.0-5.2 i$     & $39.8$ \\   
$r_{\chi} a_8^c \times 10^{5}$ & $35.0-6.2 i$     & $36.8$   &
          $44.2+3.1 i$     & $39.8$ \\   
$a_9 \times 10^{5}$   & $-936.1-13.4 i$  & $-928.4$ &
          $-953.9-24.5 i$  & $-957.3$ \\
$a_{10}^u \times 10^{5}$ & $-81.8+58.8 i$   & $-141.4$ &
           $-58.3+86.1 i$   & $-74.0$ \\
$a_{10}^c \times 10^{5} $ & $-85.2+63.5 i$   & $-141.4$ &
           $-60.3+88.8 i$   & $-74.0$ \\    
\end{tabular}
          
\vspace{0.5cm}
\caption{ The QCD coefficients $a_i^p(\pi \pi)$ at NLO and LO for the
renormalization scales at $\mu=5~ GeV$ and $\mu=2.5~GeV$, where
$r_{\chi}=\frac{2 m_{\pi}^2}{m_b (m_u+m_d)}$ }

\end{table}


\begin{figure}[tb]
\vspace*{1cm}
\centerline{\epsfig{figure=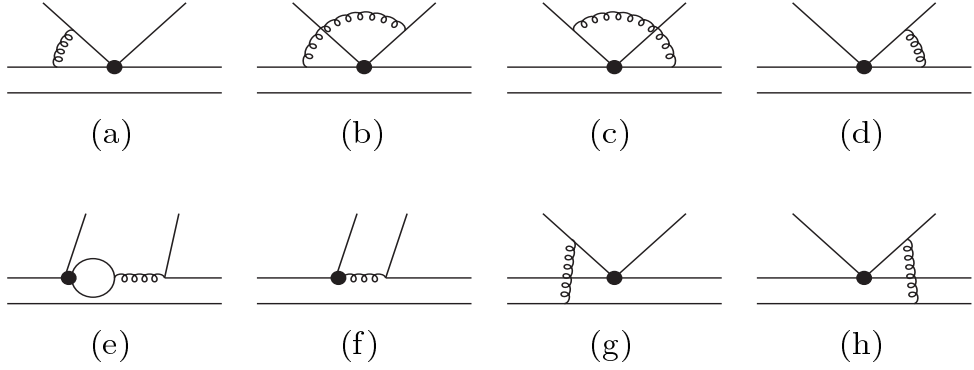,height=6cm,width=15cm,angle=0}}
\vspace*{1.cm}
\caption{Order of $\alpha_s$ corrections to hard-scattering kernels.
The upward quark lines represent the ejected quark pairs
from b quark weak decays.}
\end{figure}

\begin{figure}[tb]
\vspace*{1cm}
\centerline{\epsfig{figure=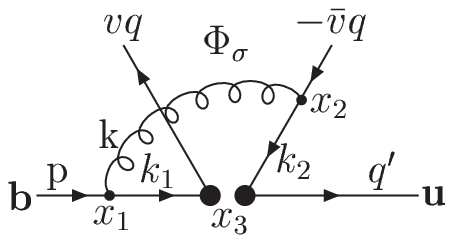,height=3cm,width=5cm,angle=0}}
\vspace*{1.cm}
\caption{}
\end{figure}

\begin{figure}[tb]
\vspace*{1cm}
\centerline{\epsfig{figure=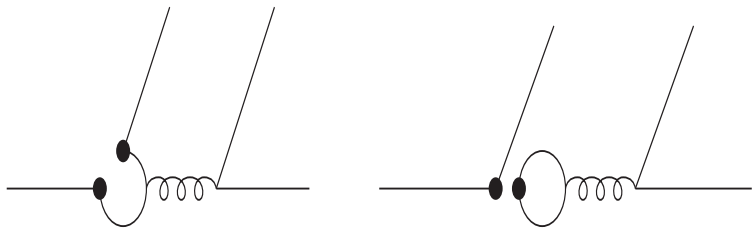,height=3cm,width=10cm,angle=0}}
\vspace*{1.cm}
\caption{ Two kinds of topology for penguin contractions. }
\end{figure}

\end{document}